\documentstyle[prb,aps]{revtex}

\begin{document}
\title{Effects of resonant tunneling in electromagnetic wave propagation through a
polariton gap\footnote{To be published in Phys. Rev. B, 1999}}
\author{Lev I. Deych$\dag$, A. Yamilov$\ddag$, and A.A. Lisyansky$\ddag$}
\address{$\dag$ Department of Physics, Seton Hall University, South Orange, NJ 07079\\
$\ddag$ Department of Physics, Queens College of CUNY, Flushing, NY 11367}
\maketitle

\begin{abstract}
We consider tunneling of electromagnetic waves through a polariton band gap
of a 1-{\it D} chain of atoms. We analytically show that a defect embedded
in the structure gives rise to the resonance transmission at the frequency
of a local polariton state associated with the defect. Numerical Monte-Carlo
simulations are used to examine properties of the electromagnetic band
arising inside the polariton gap due to finite concentration of defects.


\end{abstract}



\section{Introduction}

The resonant tunneling of electromagnetic waves through different types of
optical barriers is a fast developing area of optical physics. This effect
was first considered for photonic crystals,\cite
{Yablonovitch2,photonic} where forbidden band-gaps in the
electromagnetic spectrum form optical barriers. Macroscopic defects embedded
in the photonic crystal give rise to local photon modes,\cite%
{Yablonovitch1,Joannopoulos,Smith,Figotin,Sakoda} which induce the resonant transmission
of electromagnetic waves through the band-gaps.

A different type of photonic band gaps arises in polar dielectrics, where a
strong resonance interaction between the electromagnetic field and dipole active
internal excitations of a dielectric brings about a gap between different
branches of polaritons. Recently it was suggested that regular microscopic
impurities embedded in such a dielectric give rise to local polariton states,%
\cite{Deych1,Deych2,Podolsky} where a photon is coupled to an intrinsic
excitation of a crystal, and both these components are localized in the
vicinity of the defect.\cite{footnote}
The main peculiarity of the local polaritons is
that their electromagnetic component is bound by a {\em microscopic}
defect whose size is many order of magnitude smaller then the wavelengths
of respective photons. Another important property of these states is the
absence of a threshold for their appearance even in 3-{\em D} isotropic systems,
while for all other known local states the ``strength'' of a defect must
exceed a certain critical value before the state would split off a continous
spectrum. The reason for this peculiar behavior is a strong van Hove
singularity in the polariton density of states in the long wave region,
caused by a negative effective mass of the polariton-forming excitations of
a crystal.

The feasibility of resonant electromagnetic tunneling induced by local
polaritons, however, is not self-evident. The idea of a microscopic
defect affecting propagation of waves with macroscopic wavelength seems to
be in contradiction with common wisdom. Besides, it was found that the energy
of the electromagnetic component of local polaritons is very small compared
to the energy of its crystal counterpart. The existence of the tunneling effect was
first numerically demonstrated in Ref. \onlinecite{Deych2}, where a 1-{\em D} chain
of dipoles interacting with a scalar field imitating transverse
electromagnetic waves was considered. It was found that a single defect
embedded in such a chain results in near $100\%$ transmission at the
frequency of local polaritons through a relatively short chain ($50$ atoms). 
The frequency profile of the transmission was found to be strongly
assymetric, in contrast to  the case of electron tunneling.\cite
{electrontunneling}

In most cases (at least for a small concentration of the
transmitting centers) one-dimensional models give a reliable description of
tunneling processes, because by virtue of tunneling, a wave propagates
along a chain of resonance centers, for which a 1-{\em D} topology has the
highest probability of occurrence.\cite{Lifshitz} In our 
situation, it is also important that the local polariton states
(transmitting centers) occur without a threshold in 3-{\em D} systems as
well as in 1-{\em D} systems. This ensures that the transmission resonances
found in Ref. \onlinecite{Deych2} are not artifacts of the one-dimensional nature
of the model, and justifies a further development of the model. In the
present paper we pursue this development in two interconnected directions.
First, we present an exact analytical solution of the transmission problem
through the chain with a single defect. This solution explains the unusual
asymmetric shape of the transmission profile found in numerical calculations\cite{Deych2}
and provides insight into the phenomenon under
consideration. Second, we carry out numerical Monte-Carlo simulation of the
electromagnetic transmission through a macroscopically long chain with a
finite concentration of defects, and study the development of a
defect-induced electromagnetic pass band within the polariton band-gap. The
analytical solution of a single-defect model allows us to suggest a physical
interpretation for some of the peculiarities of the transmission found in
numerical simulations. As a by-product of our numerical results we present a
new algorithm used for the computation of the transmission. This algorithms
is based upon a blend of transfer-matrix approach with ideas of the
invariant-embedding method,\cite{IEM} and proves to be extremely stable even
deep inside the band-gap, where traditional methods would not work.

Though we consider the one-dimensional model, the results obtained are
suggestive for experimental observation of the predicted effects. Actually
the damping of the electromagnetic waves is more experimentally restrictive
than the topology of the system. We, however, discuss the effects due to
damping and come to the conclusion that the effects under discussion can be
observed in regular ionic crystals in the region of their phonon-polariton
band-gaps.

The paper is organized as follows.  The Introduction is followed by an
analytical solution of the transmission problem in a single-impurity
situation. The next section presents results of Monte-Carlo computer
simulations. The algorithm used in numerical calculations is derived and
discussed in the Appendix. The paper concludes with a discussion of the
results.

\section{ Description of the model and analytical solution of a
single-defect problem}

\subsubsection{The model}

Our system consists of a chain of atoms interacting with each other and with
a scalar ``electromagnetic'' field. Atoms are represented by their dipole
moments $P_{n}$, where the index $n$ represents the position of an atom
in the chain. Dynamics of the atoms is described within the tight-binding
approximation with an interaction between nearest neighbors only,

\begin{equation}
(\Omega _{n}^{2}-\omega ^{2})P_{n}+\Phi (P_{n+1}+P_{n-1})=\alpha E(x_{n}),
\label{dipoles}
\end{equation}
where $\Phi $ is a parameter of the interaction, and $\Omega _{n}^{2}$
represents the site energy. Impurities in the model differ from host atoms
in this parameter only, so 
\begin{equation}
\Omega _{n}^{2}=\Omega _{0}^{2}c_{n}+\Omega _{1}^{2}(1-c_{n}),
\label{site_energy}
\end{equation}
where $\Omega _{0}^{2}$ is the site energy of a host atom, $\Omega _{1}^{2}$
describes an impurity, $c_{n}$ is a random variable taking values $1$ and $0$
with probabilities $1-p$ and $p,$ respectively. Parameter $p$, therefore,
sets the concentration of the impurities in our system. This choice of the
dynamical equation corresponds to exciton-like polarization waves.
Phonon-like waves can be presented in a form that is similar to Eq. (\ref
{dipoles}) with $\Omega _{n}^{2}=\Omega
_{0}^{2}+(1-c_{n})(1-M_{def}/M_{host})\omega ^{2}$, where $M_{def}$ and $%
M_{host}$ are masses of defects and host atoms, respectively.

Polaritons in the system arise as collective excitations of dipoles
(polarization waves) coupled to the electromagnetic wave, $E(x_{n}),$ by
means of a coupling parameter $\alpha $. The electromagnetic subsystem is
described by the following equation of motion 
\begin{equation}
\frac{\omega ^{2}}{c^{2}}E(x)+\frac{d^{2}E}{dx^{2}}=-4\pi \frac{\omega ^{2}}{%
c^{2}}\sum_{n}P_{n}\delta (na-x),  \label{Maxwell}
\end{equation}
where the right hand side is the polarization density caused by atomic
dipole moments, and $c$ is the speed of light in vacuum. The coordinate $x$
in Eq. (\ref{Maxwell}) is along the chain with an interatomic distance $a$.
Eqs. (\ref{dipoles}) and (\ref{Maxwell}) present a {\it microscopic}
description of the transverse electromagnetic waves propagating along the
chain in the sense that it does not make use of the concept of the
dielectric permeability, and takes into account all modes of the field
including those with wave numbers outside of the first Brillouin band.

This approach enables us to address several general questions. A local state
is usually composed of states with all possible values of wave number $k$.
States with large $k$ cannot be considered within a macroscopic dielectric
function theory, and attempts to do so lead to divergent integrals that need
to be renormalized.\cite{Rupasov} In our approach, all expressions are well
defined, so we can check whether a contribution from large $k$ is important,
and if the long wave approximation gives reliable results. Calculation of
the integrals appearing in the 3-$D$ theory requires detailed knowledge of
the spectrum of excitations of a crystal throughout the entire Brillouin
band. This makes analytical consideration practically unfeasible. In our 1-$%
D $ model, we can carry out the calculations analytically (in a
single-impurity case) and examine the influence of different factors (and
approximations) upon the frequency of a local state and the transmission
coefficient. Using caution, the results obtained can be used to assess
approximations in 3-$D$ cases.

\subsubsection{A single impurity problem}

The equation for the frequency of the local polariton state in the 1-$D$
chain has the form similar to that derived in Ref.~[\onlinecite{Deych1}] 
\begin{equation}
1=\Delta \Omega ^{2}G(0),  \label{eigen_freq}
\end{equation}
where, however, the expression for the polariton Green's function $%
G(n-n_{0}) $ responsible for the mechanical excitation of the system can be
obtained in the explicit form 
\begin{equation}
G(n-n_{0})=\sum_{k}{\displaystyle}\frac{\cos (ak)-\cos (\frac{a\omega }{c})}{%
\left[ \omega ^{2}-\Omega _{0}^{2}-2\Phi \cos \left( ka\right) \right] \left[
\cos (ak)-\cos (\frac{a\omega }{c})\right] -\displaystyle{\frac{2\pi \alpha
\omega }{c}}\sin (\frac{a\omega }{c})}\exp \left[ ik\left( n-n_{0}\right) a%
\right] .  \label{Greenfunction}
\end{equation}
If one neglects the term responsible for the coupling to the electromagnetic
field, the Green's function $G(n-n_{0})$ is reduced to that of the pure
atomic system. This fact reflects the nature of the defect in our model: it
only disturbs the mechanical (not related to the interaction with the field)
properties of the system. A solution of Eq. (\ref{eigen_freq}) can be
real-valued only if it falls into the gap between the upper and lower
polariton branches. This gap exists if the parameter $\Phi $ in the
dispersion equation of the polariton wave is positive, and the effective
mass of the excitations in the long wave limit is, therefore, negative.

The diagonal element, $G(0)$, of Green's function (\ref{Greenfunction}) can
be calculated exactly. The dispersion equation (\ref{eigen_freq}) than takes
the following form: 
\begin{equation}
1=\Delta \Omega ^{2}\frac{1}{2\Phi D(\omega )}\left[ \frac{\cos (\frac{%
a\omega }{c})-Q_{2}(\omega )}{\sqrt{Q_{2}^{2}(\omega )-1}}-\frac{\cos (\frac{%
a\omega }{c})-Q_{1}(\omega )}{\sqrt{Q_{1}^{2}(\omega )-1}}\right] ,
\label{eigen_freq_int}
\end{equation}
where $Q_{1,2}(\omega )$, 
\begin{eqnarray}
Q_{1,2}(\omega ) &=&\frac{1}{2}\left[ \cos (\frac{a\omega }{c})+\frac{\omega
^{2}-\Omega _{0}^{2}}{2\Phi }\right] \pm \frac{1}{2}D(\omega ),
\label{determinant} \\
D(\omega ) &=&\sqrt{\left[ \cos (\frac{a\omega }{c})-\frac{\omega
^{2}-\Omega _{0}^{2}}{2\Phi }\right] ^{2}-\frac{4\pi \alpha \omega }{\Phi c}%
\sin (\frac{a\omega }{c})}
\end{eqnarray}
give the poles of the integrand in Eq. (\ref{Greenfunction}). The bottom of
the polariton gap is determined by the condition $D(\omega )=0,$ yielding in
the long wave limit, $a\omega /c\ll 1$, for the corresponding frequency, $%
\omega _{l}$, 
\begin{equation}
\omega _{l}^{2}\simeq \tilde{\Omega}_{0}^{2}-2\tilde{\Omega}_{0}^{2}d\frac{%
\sqrt{\Phi }a}{c},
\end{equation}
where we introduce parameters $d^{2}=4\pi \alpha /a$, $\tilde{\Omega}_{0}^{2}
$ $=\Omega _{0}^{2}+2\Phi $, and take into account that the band width of
the polarization waves, $\Phi $, obeys the inequality $\sqrt{\Phi }a/c\ll 1$%
. The last term in this expression is the correction to the bottom of the
polariton gap due to the interaction with the transverse electromagnetic
field. Usually this correction is small, but it has an important
theoretical, and, in the case of strong enough spatial dispersion and
oscillator strength, practical significance.\cite{Deych1} Because of this
correction the polariton gap starts at a frequency, when the determinant $%
D(\omega )$ becomes imaginary, but functions $Q_{1,2}(\omega )$ are still
less than $1$. This leads to the divergence of the right-hand side of Eq. (%
\ref{eigen_freq_int}) as $\omega $ approaches $\omega _{l}$, and, hence, to
the absence of a threshold for the solution of this equation. This
divergence is not a 1-{\em D} effect since the same behavior is also found
in 3-{\em D} isotropic model.\cite{Deych1,Podolsky} An asymptotic form for
Eq. (\ref{eigen_freq_int}) when $\omega $ $\longrightarrow $ $\omega _{l}$
in the 1-{\em D} case reads 
\begin{equation}
\sqrt{\omega ^{2}-\omega _{l}^{2}}\sim \frac{\Delta \Omega ^{2}}{\sqrt{\Phi }%
},
\end{equation}
and differs from the 3-{\em D} case by the factor of $\left( d\omega
_{l}a)/(c\sqrt{\Phi }\right) $.  The upper boundary of the gap, $\omega
_{up}$, is determined by the condition $Q_{1}(\omega )=0$, leading to 
\begin{equation}
\omega _{up}^{2}=\tilde{\Omega}_{0}^{2}+d^{2},  \label{wup}
\end{equation}
Eq. (\ref{eigen_freq_int}) also has a singularity as $\omega $ $%
\longrightarrow $ $\omega _{up}$, but this singularity is exclusively caused
by the 1-{\em D} nature of the system. We will discuss local states that are
not too close to the upper boundary in order to avoid manifestations of
purely 1-{\em D} effects.

For frequencies deeper inside the gap, Eq. (\ref{eigen_freq}) can be
simplified in the approximation of small spatial dispersion, $\sqrt{\Phi }%
a/c\ll 1,$ to yield 
\begin{equation}
\omega ^{2}=\tilde{\Omega}_{1}^{2}-\Delta \Omega ^{2}\left[ 1-\sqrt{\frac{%
\omega ^{2}-\tilde{\Omega}_{0}^{2}}{\omega ^{2}-\tilde{\Omega}_{0}^{2}+4\Phi 
}}\right] -d^{2}\frac{\omega a}{2c}\frac{\Delta \Omega ^{2}}{\sqrt{\left(
\omega ^{2}-\tilde{\Omega}_{0}^{2}\right) \left( \tilde{\Omega}%
_{0}^{2}+d^{2}-\omega ^{2}\right) }},  \label{nodispersion}
\end{equation}
where $\tilde{\Omega}_{1}^{2}=\Omega _{1}^{2}+2\Phi $ is a fundamental ($%
k=0) $ frequency of a chain composed of impurity atoms only. Two other terms
in Eq. (\ref{nodispersion}) present corrections to this frequency due to the
spatial dispersion and the interaction with the electromagnetic field
respectively. One can see that both corrections have the same sign and shift
the local frequency into the region between $\tilde{\Omega}_{0}^{2}$ and $%
\tilde{\Omega}_{1}^{2}$. As we see below, this fact is significant for
the transport properties of the chain.

Transmission through the system can be considered in the framework of the
transfer matrix approach. This method was adapted for the particular case of
the system under consideration in Ref. \onlinecite{Deych2}. The state of the system
is described by the vector $v_{n},$ with components $P_{n}$, $P_{n+1}$, $E_{n}$, 
$E_{n}^{\prime }/{k}_{\omega }$), which obeys to the following difference
equation 
\begin{equation}
v_{n+1}=T_{n}v_{n}.  \label{EP}
\end{equation}
The transfer matrix $T_{n}$ describes the propagation of the vector between
adjacent sites: 
\begin{equation}
T_{n}=\left( 
\begin{array}{cccc}
{0} & {1} & {0} & {0} \\ 
{-1} & {{-\displaystyle\frac{\Omega _{n}^{2}-\omega ^{2}}{\Phi }}} & %
\displaystyle{{\frac{\alpha }{\Phi }\cos {ka}}} & \displaystyle{{{\frac{%
\alpha }{\Phi }}\sin {ka}}} \\ 
{0} & {0} & {\cos {ka}} & {\sin {ka}} \\ 
{0} & {-4\pi k} & {-\sin {ka}} & {\cos {ka}}
\end{array}
\right) .  \label{T}
\end{equation}

Analytical calculation of the transmission coefficient in the situation
considered is not feasible even in the case of a single impurity because the
algebra is too cumbersome. The problem, however, can be 
simplified considerably if one neglects the spatial dispersion of the polarization waves.
In this case the $T$-matrix can be reduced to a $2\times 2$ matrix of the
following form 
\begin{equation}
\tau _{n}=\left( 
\begin{array}{cc}
\cos ka & \sin ka \\ 
-\sin ka+\beta _{n}\cos ka & \cos ka+\beta _{n}\sin ka
\end{array}
\right) ,  \label{T reduced}
\end{equation}
where the parameter $\beta _{n}$ 
\begin{equation}
\beta _{n}=\frac{4\pi \alpha \omega }{c\left( \omega ^{2}-\Omega
_{n}^{2}\right) },  \nonumber
\end{equation}
represents the polarizability of the $n$-th atom due to its vibrational
motion. In this case the complex transmission coefficient, $t$, can 
be easily expressed in terms of the elements of the resulting transfer matrix, $%
T^{(N)}=\prod_{1}^{N}\tau _{n}$, 
\begin{equation}
t=\frac{2}{\left( T_{11}^{(N)}+T_{22}^{(N)}\right) -i\left(
T_{12}^{(N)}-T_{21}^{(N)}\right) }e^{-ikL}.  \label{transmcoeff}
\end{equation}
The problem is, therefore, reduced to the calculation of $T^{(N)}$. In the
case of a single impurity, the product of the transfer matrices, $\tau $,
can be presented in the following form 
\begin{equation}
T^{(N)}={\tau }^{N-n_{0}}\times \tau _{def}\times {\tau }^{n_{0}-1},
\label{iproduct}
\end{equation}
where the matrix $\tau _{def}$ describes the impurity atom with $\Omega
_{n}=\Omega _{1}$. The matrix product in Eq. (\ref{iproduct}) is
conveniently calculated in the basis, where the matrix $\tau $ is diagonal.
After some cumbersome algebra, one obtains for the complex transmission
coefficient: 
\begin{equation}
t=\frac{2e^{ikL}\exp \left( -\kappa L\right) }{\left[ 1-\frac{i}{\sqrt{R}}%
\left( 2-\beta \cot ka\right) \right] \left[ \left( 1+\varepsilon \right) 
\right] +2i\exp \left( -\kappa L\right) \Gamma \cosh \left[ \kappa
a(N-2n_{0}+1\right] }.  \label{transmission}
\end{equation}
where $R=\beta ^{2}+4\beta \cot (ak)-4$, $\Gamma =\varepsilon \beta /(\sin
(ka)\sqrt{R})$, $\kappa $ is the imaginary wave number of the evanescent
electromagnetic excitations, which determines the inverse localization length of
the local state, and $\varepsilon =\left( \beta _{def}-\beta \right) /2\sqrt{%
R}$. The last parameter describes the difference between host atoms and the
impurity, and is equal to 
\begin{equation}
\varepsilon =\frac{2\pi \alpha }{c\sqrt{R}}\omega \frac{\left( \Omega
_{1}^{2}-\Omega _{0}^{2}\right) }{\left( \omega ^{2}-\Omega _{0}^{2}\right)
\left( \omega ^{2}-\Omega _{1}^{2}\right) }.  \label{epsilon}
\end{equation}
We have also neglected here a contribution from the second eigenvalue of the
transfer matrix, which is proportional to $\exp (-2\kappa L)$, and is
exponentially small for sufficiently long chains. For $\varepsilon =0,$ Eq. (\ref{transmission})
gives the transmission coefficient, $t_{0},$ of the pure system, 
\begin{equation}
t_{0}=\frac{2e^{ikL}\exp \left( -\kappa L\right) }{1-\frac{i}{\sqrt{R}}%
\left( 2-\beta \cot ka\right) },  \label{tpure}
\end{equation}
exhibiting a regular exponential decay. At the lower boundary of the
polariton gap, $\Omega _{0}$, parameters $\beta $ and $\kappa $ diverge,
leading to vanishing transmission at the gap edge regardless the length of
the chain. It is instructive to rewrite Eq. (\ref{transmission}) in terms of 
$t_{0}$: 
\begin{equation}
t=\frac{t_{0}}{\left( 1+\varepsilon \right) +i\exp \left( -ikL\right) \Gamma
t_{0}\cosh \left[ \kappa a(N-2n_{0}+1\right] }  \label{transmission1}
\end{equation}
This expression describes the resonance tunneling of the electromagnetic
waves through the chain with the defect. The resonance occurs when 
\begin{equation}
1+\varepsilon =0,  \label{res_freq}
\end{equation}
the transmission in this case becomes independent of the system size.
Substituting the definition of the parameter $\varepsilon $ given by Eq. (%
\ref{epsilon}) into Eq. (\ref{res_freq}), one arrives at an equation
identical to Eq. (\ref{nodispersion}) for the frequency of the local
polariton state with the parameter of the spatial dispersion, $\Phi$, being
set to zero. The transmission takes a maximum value when the defect is
placed in the middle of the chain, $N-2n_{0}+1=0$, and in this case 
\begin{equation}
|t_{\max }|^{2}=\frac{1}{\Gamma ^{2}}\leq 1.  \label{tmax}
\end{equation}
The width of the resonance is proportional to $\Gamma t_{0}$ and decreases
exponentially with an increase of the system's size. In the long wave limit, $%
ak\ll 1,$ Eq. (\ref{tmax}) can be rewritten in the following form 
\begin{equation}
|t_{\max }|^{2}=1-\left( 1-2\frac{\omega _{r}^{2}-\Omega _{0}^{2}}{d^{2}}%
\right) ,  \label{tmaxlongwave}
\end{equation}
where $\omega _{r}$ is the resonance frequency satisfying Eq. (\ref{res_freq}%
). It is interesting to note that the transmission coefficient becomes
exactly equal to one if the resonance frequency happens to occur exactly in
the center of the polariton gap. This fact has a simple physical meaning.
For $\omega _{r}^{2}=\Omega _{0}^{2}+d^{2}/2$ the inverse localization
length $\kappa $ becomes equal to the wave number $\omega _{r}/c$ of the
incoming radiation. Owing to this fact, the field and its derivative inside
the chain exactly match the field and the derivative of the incoming field
as though the optical properties of the chain are identical to those in
vacuum. Consequently, the field propagates through the chain without
reflection.

Having solved the transmission problem we can find the magnitude of the
field inside the chain in terms of the incident amplitude $E_{in}$ at the
resonance frequency. Spatial distribution of the field in the local
polariton state can be found to have the form $E=E_{d}\exp \left(
-|n-n_{0}|\kappa a\right)$. Matching this expression with the outcoming
field equal to $E_{in}t\exp (ikL)$ one has for the field amplitude at the
defect atom, $E_{d}$, 
\begin{equation}
E_{d}=E_{in}t\exp (-ikL)\exp \left[ (N-n_{0})\kappa a\right] .
\label{defect _field}
\end{equation}
For $|t|$ being of the order of one in the resonance this expression
describes the drastic exponential enhancement of the incident amplitude at
the defect side due to the effect of the resonance tunneling.

Equations (\ref{transmission1}) and (\ref{tmax}) demonstrate that the
resonance tunneling via local polariton states is remarkably different from
other types of resonance tunneling phenomena, such as electron tunneling via
an impurity state,\cite{electrontunneling} or through a double barrier. The
most important fact is that the frequency profile of the resonance does not
have the typical symmetric Laurentian shape. At $\omega =\Omega _{1}$ the
parameter $\varepsilon $ diverges causing the transmission to vanish. At the
same time the resonance frequency $\omega _{r}$ is very close to $\Omega
_{1} $ as it follows from Eq. (\ref{nodispersion}). This results in strongly
asymmetric frequency dependence of the transmission, which is skewed toward
lower frequencies.

The transmission vanishes precisely at two frequencies: at the low frequency
band edge $\Omega _{0}$ and at the frequency $\Omega _{1}$ associated with
the vibrational motion of the defect atom. At the same time, the behavior of
the transmission coefficient in the vicinities of these two frequencies is
essentially different: at the band edge it is $(\omega ^{2}-\Omega
_{0}^{2})^{2}\exp \left( -1/\sqrt{\omega ^{2}-\Omega _{0}^{2}}\right) $
while at the defect frequency the transmission goes to zero as $(\omega
^{2}-\Omega _{1}^{2})^{2}$. These facts can be used to predict several
effects that would occur with the increase of the concentration of the
defects. First, one can note that with the increase of concentration of the
impurities frequency $\Omega _{1}$ becomes eventually the boundary of the
new polariton gap when all the original host atoms will be replaced by the
defects atoms. One can conclude then that the zero of the transmission at $%
\Omega _{1}$ instead of being washed out by the disorder, would actually
become more singular. More exactly one should expect that the frequency
dependence of the transmission in the vicinity of $\Omega _{1}$ will exhibit
a crossover from the simple power decrease to the behavior with exponential
singularity associated with the band edge. Second, if one takes into account
such factors as spatial dispersion or damping, which prevent transmission
from exact vanishing, one should expect that the above mentioned crossover
to the more singular behavior would manifest itself in the form of
substantial decrease of the transmission in the vicinity of $\Omega _{1}$
with an increase of the concentration. Numerical calculations discussed in
the next section of the paper show that this effect does take place even at
rather small concentration of the defects.

Resonance tunneling is very sensitive to the presence of relaxation, which
phenomenologically can be accounted for by adding $2i\gamma \omega $ to the
denominator of the polarizability $\beta $, where $\gamma $ is an effective
relaxation parameter. This will make the parameter $\epsilon $ complex
valued, leading to two important consequences. First, the resonance
condition becomes $Re(\varepsilon )=-1$, and it can be fulfilled only if the
relaxation is small enough. Second, the imaginary part of $\varepsilon $
will prevent the exponential factor $t_{0}$ in Eq. (\ref{transmission1})
from canceling out at the resonance. This restricts the length of the system
in which the resonance can occur and limit the enhancement of the field at
the defect. These restrictions though are not specific for the system under
consideration and affect experimental manifestation of any type of resonant
tunneling phenomenon.

Since we only concern with a frequency region in the vicinity of $\Omega
_{1} $, real, $\varepsilon _{1}$, and imaginary, $\varepsilon _{2}$, parts
of $\varepsilon $ can be approximately found as 
\begin{equation}
\varepsilon _{1}\simeq d^{2}\frac{\Omega _{1}a}{2c}\sqrt{\frac{\Delta \Omega
^{2}}{d^{2}-\Delta \Omega ^{2}}}\frac{\omega ^{2}-\Omega _{1}^{2}}{\left(
\omega ^{2}-\Omega _{1}^{2}\right) ^{2}+4\gamma ^{2}\omega ^{2}},
\label{Re_eps}
\end{equation}
\begin{equation}
\varepsilon _{2}\simeq \frac{2\gamma \omega }{\omega ^{2}-\Omega _{1}^{2}}%
\varepsilon _{1}.  \label{Im_eps}
\end{equation}
It follows from Eq. (\ref{Re_eps}) that the resonance occurs only if $%
(4\gamma c)/(ad^{2})<1$. This inequality has a simple physical meaning: it
ensures that the distance between the resonance frequency, $\omega _{r}$,
and $\Omega _{1}$, where the transmission goes to zero, is greater than the
relaxation parameter, $\gamma $. This is a rather strict condition that can
only be satisfied for high frequency oscillations with large oscillator
strength in crystals with large interatomic spacing, $a$. The spatial
dispersion, however, makes conditions for the resonant tunneling much less restrictive. In order to
estimate the effect of the dissipation in the presence of the spatial
dispersion one can rely upon Eq. (\ref{transmission1}) assuming that the
dispersion only modifies the parameter $\varepsilon $, but does not effect
the general expression for the transmission. This assumption is justified by the numerical results of Ref. \onlinecite
{Deych2} and the present paper, which show that the transmission properties in the presence of the spatial dispersion do
not differ significantly from the analytical calculations performed for the chain of noninteracting dipoles. According to Eq. (\ref
{nodispersion}), the inter-atomic interaction moves the resonance frequency
further away from $\Omega _{1}$ undermining the influence of the damping and
leading to a weaker inequality: $(\gamma \Omega _{1})/\Phi <1$. This
condition can be easily fulfilled, even for phonons with a relatively small
negative spatial dispersion. For the imaginary part $\varepsilon _{2}$ at
the resonance one can obtain from Eq. (\ref{Im_eps}) the following estimate 
\begin{equation}
\varepsilon _{2}\sim \min [(4\gamma c)/(ad^{2}),(\gamma \Omega _{1})/\Phi ].
\end{equation}
The requirement that $\varepsilon _{2}$ be much smaller than $t_{0}$ leads
to the following restriction for the length of the system $L\ll (1/\kappa
)\mid \ln [\varepsilon _{2}]\mid $, with $\varepsilon _{2}$ given above. The
maximum value of the field at the defect site attainable for the defect
located in the center of the chain is then found as $|E_{d}|\sim |E_{in}||t|/%
\sqrt{\varepsilon _{2}}$.

\section{One-dimensional dipole chain with finite concentration of impurities}

In this section we present results of numerical Monte-Carlo simulations of
the transport properties of the system under consideration in the case of
randomly distributed identical defects. If spatial dispersion is taken into
account the regular Maxwell boundary conditions must be complemented by
additional boundary conditions regulating the behavior of polarization $P$
at the ends of the chain. In our previous paper\cite{Deych2} we calculated
the transmission for two types of boundary conditions: $P_{0}=P_{N}=0$,
which corresponds to the fixed ends of the chain, and $P_{0}=P_{1}$, $%
P_{N-1}=P_{N}$, which corresponds to the relaxed ends. We reported in Ref. 
\onlinecite{Deych2}  that the transmission is very sensitive to the boundary
conditions with fixed ends being much more favorable for the resonance. Our
present numerical results obtained with an improved numerical procedure and
the analytical calculations do not confirm this dependence of the resonant
tunneling upon the boundary conditions. In the case of a single impurity 
we find that for both types of the boundary conditions the transmission
demonstrates sharp resonance similar to that found in Ref. \onlinecite{Deych2} for
fixed ends. Similarly, for a finite concentration of impurities we did not find any considerable differences in the transmission for both types of boundary conditions. We conclude that
the actual form of the boundary conditions is not
significant for the resonant tunneling. 

The transfer matrix, Eq. (\ref{EP}), along with the
definition of the transfer matrix, Eq. (\ref{T}), and the boundary
conditions chosen in the form of fixed terminal points, provides a basis for our computations. However, it turns out that
straightforward use of Eq. (\ref{EP}) in the gap region is not possible
because of underflow errors arising when one pair of eigenvalues of the
transfer matrix becomes exponentially greater than the second one. In order
to overcome this problem we develop a new computational approach based upon
the blend of the transfer-matrix method with the invariant embedding ideas.
The central element of the method is a $4\times 4$ matrix $S(N)$ that
depends upon the system size $N$. The complex transmission coefficient, $t,$
is expressed in terms of the elements of this matrix as 
\begin{equation}
t=2\exp (-ikL)(S_{11}+S_{12}).  \label{tfromS}
\end{equation}
The matrix $S(N)$ is determined by the following nonlinear recursion 
\begin{equation}
S(N+1)=T_{N}\times \Xi (N)\times S(N),  \label{S}
\end{equation}
where matrix $\Xi (N)$ is given by 
\begin{equation}
\Xi (N)=\{I-S(N)\times H\times \left[ I-T(N)\right] \}^{-1}.  \label{F}
\end{equation}
The initial condition to Eq. (\ref{S}) is given by 
\begin{equation}
S(0)=(G+H)^{-1},  \label{S_init}
\end{equation}
where matrices $G$ and $H$ are specified by the boundary conditions. The
derivation of the Eqs. (\ref{tfromS})-(\ref{S_init}) and more detailed
discussion of the method is given in the Appendix to the paper. The test of
the algorithm based upon recursion formula (\ref{S}) proves the method to
provide accurate results for transmission coefficients as small as $10^{-15}$.

In our simulations we fix the concentration of the defects and randomly
distribute them among the host atoms. The total number of atoms in the chain
is also fixed; the results presented below are obtained for a chain
consisting of $1000$ atoms. For the chosen defect frequency, $\Omega
_{1}\simeq 1.354\Omega _{0}$, the localization length of the local polariton
state, $l_{ind},$is approximately equal to $150$ interatomic distances. The
transmission coefficient is found to be extremely sensitive to a particular
arrangements of defects in a realization exhibiting strong fluctuations from
one realization to another. Therefore, in order to reveal the general
features of the transmission independent of particular positions of defects,
we average the transmission over $1000$ different realizations. We have also
calculated the averaged Lyapunov exponent (the inverse localization length, $%
l_{chain}$, characterizing transport through the entire chain) to verify
that the averaged transmission reveals a reliable information about the
transport properties of the system.

The results of the computations are presented in the figures below. Figs. 1-3
show an evolution of the transmission with the increase of the
concentration of the impurities. In Fig. 1 one can see the change of the
transport properties at small concentrations up to 1\%. The curve labeled
(1) shows, basically, the single impurity behavior averaged over random
positions of the defect. With an increase of the concentration there is a
greater probability for two (or more) defects to form a cluster resulting in
splitting a single resonance frequency in two or more frequencies. The
double peak structure of the curves (2) and (3) reflects these cluster
effects. With the further increase of the concentration the clusters' sizes
grow on average leading to multiple resonances with distances between
adjacent resonance frequencies being too small to be distinguished. Curve
(5) in Fig. 1 reflects this transformation, which marks a transition between
individual tunneling resonances and the defect induced band. The
concentrations in this transition region is such that an average distance
between the defects is equal to the localization length of the individual
local states, $l_{ind}$. The collective localization length at the frequency
of the transmission peak, $l_{chain}^{\max }$, becomes equal to the length
of the chain at approximately the same concentration that allows to suggest
a simple relationships between the two lengths: $l_{chain}^{\max }=cl_{ind}$,
where $c$ stands for the concentration. The numerical results presented in
Fig. 4 clearly demonstrate this linear concentration dependence of $%
l_{chain}^{\max }$ at small concentrations. For larger concentrations one
can see from Figs. 2 and 3 that the peak of the transmission coefficient
developes into a broad structure. This marks further development of the
defect pass band. Curves in Fig. 2 show the transmission coefficient at
intermediate concentrations, where localization length, $l_{chain},$ is
bigger than the length of the system only in a small frequency region around
the maximum of the transmission, and Fig. 3 presents well developed pass
band with multipeak structure resulting from geometrical resonances at the
boundaries of the system.

This figures reveal an important feature of the defect polariton band: its
right edge does not move with increase of the concentration. The frequency
of this boundary is exactly equal to the defect frequency, $\Omega _{1},$
(which is normalized by $\Omega _{0}$ in the figures), and the entire band
is developing to the left of $\Omega _{1}$ in complete agreement with the
arguments based upon analytical solution of the single-impurity problem.
Moreover, the magnitude of the transmission in the vicinity of $\Omega _{1}$%
decreases with increase of the concentration also in agreement with our
remarks at the end of the previous section. Fig. 5 presents the inverse
localization length, $l_{chain},$ normalized by the length of the chain for
three different concentrations. It can be seen that $l_{chain}^{-1}($ $%
\Omega _{1})$ significantly grows with increase of the concentration,
reaching the value of approximately $17/L$ at the concentration as small as $%
3\%$. Such a small localization length corresponds to the transmission of
the order of magnitude of $10^{-17}$, which is practically zero in our
computation. Further increase of the concentration does not change the
minimum localization length. These results present an interesting example of
the defects building up a boundary of the new forbidden gap.

From this figure it is also seen the development of the pass band to the
left of $\Omega _{1}$ presented above in Figs. 1-3, but at a larger scale.
We can not distinguish here the details of the frequency dependence, but the
transition from the single resonance behavior to the pass band, marked by
the significant flattening of the curve, is clear.

The last Fig. 6 presents the concentration dependence of the semiwidth, $%
\delta _{\omega },$ of the defect band. The semiwidth is defined as the
difference between frequency of the maximum transmission and the right edge
of the band. One can see that all the point form a smooth line with no
indication of a change of the dependence with the transition between
different transport regimes. Attempts to fit this curve showed that it is
excellently fitted by the power law $\delta _{\omega }\propto c^{\nu }$with $%
\nu \simeq 0.8$ in all studied concentration range. The reason for this
behavior and why it is insensitive to the change of the character of the
transport requires further study.

\section{Conclusion}

In this paper, we considered one-dimensional resonance tunneling of scalar
``electromagnetic waves'' through an optical barrier caused by a polariton
gap. The tunneling is mediated by local polariton states\ arising due to
defect atoms embedded in an otherwise ideal periodic chain. We also
numerically studied how a defect-induced propagating band emerges from these
resonances when the concentration of defects increases.

It is important to emphasize the difference between the situation considered
in our paper and other types of tunneling phenomena discussed in the
literature. The tunneling of electromagnetic waves through photonic crystals
and electron tunneling, despite all the difference between these phenomena,
share one common feature. In both cases, the resonance occurs due to defects
that have dimensions comparable with wavelengths of the respective
excitations (electrons interact with atomic impurities, and long wave
electromagnetic waves interact with macroscopic distortions of the photonic
crystals). In our case the wavelength of the propagating excitations is many
orders of magnitude greater than dimensions of the atomic defects
responsible for the resonance. The physical reason for such an unusual
behavior lies in the nature of local polaritons. These states are formed  owing to the presence of internal polariton-forming
excitations. The spatial extent of these states is much larger than the geometrical dimensions of atomic defects and is 
comparable to the wavelength of the incident radiation.

We  presented an exact analytical solution of the tunneling of
electromagnetic waves through a chain of noninteracting atoms with a single
defect. This solution provides insight into the nature of the
phenomenon under considerarion and allows one to obtain an explicit
expression for the magnitude of the electromagnetic field at the defect
site. The expression derived demonstrates that the field is strongly
enhanced at the resonance with its magnitude growing exponentially with an
increase of the length of the system. This effect is an electromagnetic
analogue of the charge accumulation in the case of electron tunneling, where
it is known to cause interesting nonlinear phenomena.\cite%
{Penley,Azbel,Goldman-1,Goldman-2,Goldman-3}

An analytical solution of the single-defect problem allowed us to make
predictions regarding the transport properties of the system with multiple
randomly located defects. The most interesting of these is that the
dynamical frequency of the defects, $\Omega _{1},$ sets a high frequency
boundary for the defect induced pass band, which does not move with
increasing concentration of defects. Numerical Monte-Carlo simulations
confirmed this assumption and showed that the direct interaction between
the atoms (spatial dispersion) does not affect resonance tunneling
considerably, though it adds new interesting features to it. One of them is
the behavior of the transmission in the vicinity of $\Omega _{1}$. In
absence of the spatial dispersion, the transmission at this point is exactly
equal to zero, and remains small when the interaction is taken into account.
The interesting fact revealed by the numerical analysis is that the
transmission at $\Omega _{1}$ decreases with an increase in the
concentration of the defects and nearly approaches zero at concentrations as
small as $3\%$. This fact can be understood in light of the transfer matrix
approach: if the frequency $\Omega _{1}$ corresponds to the eigenvalue of the defect's
transfer matrix, which significantly differs from one, the transmission will
diminish strongly each time the wave encounters a defect site, regardless
the order in which the defects are located. Numerical results also
demonstrated a transition between two transport regimes: one associated with
resonance tunneling and the other occurring when the resonances spatially
overlap and a pass band of extended states emerges. The transition occurs
when the average distance between the defects becomes equal to the
localization length of the single local state. At the same time the
collective localization length at the peak transmission frequency,
characterizing the transport properties of the entire chain, becomes equal
to the total length of the system. This result assumes the linear dependence
of this collective localization length upon concentration, which we directly
confirm for small concentrations. Numerical results also showed that the
width of the resonance, which develops into a pass band with an increase in
concentration, does not manifest any transformation when the character of
transport changes. The concentration dependence of the width was found to be
extremely well described by a power law with an exponent approximately equal
to $0.8$. The nature of this behavior awaits an explanation.

\appendix

\section{Invariant Embedding Algorithm for the Transfer-Matrix Equation}

In this Appendix we develop invariant embedding approach to transfer-matrix
equations of a general form and deduce Eqs. (\ref{tfromS})-(\ref{S_init})
used for Monte-Carlo calculations in our paper. Consider a typical
difference equation of the transfer-matrix method, 
\begin{equation}
u_{n+1}=T_{n}u_{n},  \label{Eq1}
\end{equation}
with boundary conditions of a general form: 
\begin{equation}
Gu_{0}+Hu_{N}=v.  \label{Eq2}
\end{equation}
Here $u_{n}$ is a vector of an appropriate dimension that characterizes the
state of the system at the $n$th site, $T_{n}$ is a respective
transfer-matrix; $G$ and $H$ are matrices of the same dimension as the
transfer-matrix, together a with the vector $v$ they specify boundary
conditions at the left and right boundaries of the system (cites $0$ and $N$
respectively). The regular Maxwell boundary conditions and the fixed ends
boundary condition for polarization can be presented in the form Eq. (\ref
{Eq2}) with the following matrices $G$,$H$, and vector $v$ 
\begin{equation}
G=\left( 
\begin{array}{cc}
\begin{array}{cc}
1 & -i \\ 
1 & -i
\end{array}
& 
\begin{array}{cc}
0 & 0 \\ 
0 & 0
\end{array}
\\ 
\begin{array}{cc}
0 & 0 \\ 
0 & 0
\end{array}
& 
\begin{array}{cc}
1 & 0 \\ 
1 & 0
\end{array}
\end{array}
\right) ,\qquad H=\left( 
\begin{array}{cc}
\begin{array}{cc}
1 & i \\ 
-1 & -i
\end{array}
& 
\begin{array}{cc}
0 & 0 \\ 
0 & 0
\end{array}
\\ 
\begin{array}{cc}
0 & 0 \\ 
0 & 0
\end{array}
& 
\begin{array}{cc}
0 & 1 \\ 
0 & -1
\end{array}
\end{array}
\right) ,\qquad v=\left( 
\begin{array}{c}
2 \\ 
2 \\ 
0 \\ 
0
\end{array}
\right) .  \label{BC}
\end{equation}
These matrices are singular, but one should not worry about this, because we
will only need to invert their sum, which has a non-zero determinant. In
accord with the ideas of the invariant embedding method\cite{IEM} we
consider the dynamic vector, $u_{n},$ as a function of the current site $n$,
the length of the system $N$, and the boundary vector $v$: 
\begin{equation}
u_{n}\equiv u(n,N,v)\equiv S(n,N)v  \label{Eq3}
\end{equation}
In the last equation we use the linear nature of Eq. (\ref{Eq1}) in order to
separate out the dependence upon the vector $v$. Substituting Eq. (\ref{Eq3}%
) into Eqs. (\ref{Eq1}) and (\ref{Eq2}) we have the dynamical equation and
boundary conditions for the matrix $S:$%
\begin{equation}
S(n+1,N)=T_{n}\times S(n,N),  \label{Eq4a}
\end{equation}
\begin{equation}
G\times S(0,N)+H\times S(N,N)=I,  \label{Eq4b}
\end{equation}
where $I$ is a unit matrix. The matrix $S(n,N+1)$, which describes the
system with one additional scatterer, obviously satisfies the same equation (%
\ref{Eq4a}) as $S(n,N)$. Relying again upon the linearity of Eq. (\ref{Eq4a}%
) we conclude that $S(n,N)$ and $S(n,N+1)$ can only differ by a constant
(independent of $n$) matrix factor $\Lambda (N)$. 
\begin{equation}
S(n,N+1)=S(n,N)\times \Lambda (N).  \label{Eq5}
\end{equation}
In order to find $\Lambda (N)$ we first substitute Eq. (\ref{Eq5}) into
boundary conditions Eq. (\ref{Eq4b}) which yield 
\begin{equation}
\Lambda (N)=G\times S(0,N+1)+H\times S(N,N+1).  \label{Eq6}
\end{equation}
Boundary conditions Eq. (\ref{Eq4b}) do not change if $N$ is replaced by $%
N+1 $, therefore we can write down that 
\begin{equation}
G\times S(0,N+1)=I-H\times S(N+1,N+1).  \nonumber
\end{equation}
Substituting this expression into Eq. ( \ref{Eq6}) we have for the matrix $%
\Lambda (N):$%
\begin{equation}
\Lambda (N)=I+H\times \lbrack S(N,N+1)-S(N+1,N+1)].  \label{Eq7}
\end{equation}
The quantity $S(N+1,N+1)$ can be eliminated from this equation by means of
Eq. (\ref{Eq1}): $S(N+1,N+1)=T_{N}$ $S(N,N+1)$, and we have for $\Lambda (N)$%
\begin{equation}
\Lambda (N)=I+H\times \lbrack I-T(N)]\times S(N,N+1).  \label{Eq8}
\end{equation}
Substituting this formula into Eq. (\ref{Eq5}) we obtain the equation that
governs the evolution of the matrix $S(n,N)$ with the change of the
parameter $N$:

\begin{equation}
S(n,N+1)=S(n,N)+S(n,N)\times H\times \lbrack I-T(N)]\times S(N,N+1).
\label{Eq9}
\end{equation}
This equation, however, is not closed because of an unknown matrix $S(N,N+1)$%
. This matrix can be found by setting $n=N$ in Eq. (\ref{Eq9}): 
\begin{equation}
S(N,N+1)=\{I-S(N,N)\times H\times \lbrack I-T(N)]\}^{-1}S(N,N).  \nonumber
\end{equation}
Introducing notation 
\begin{equation}
\Xi (N)=\{I-S(N,N)\times H\times \lbrack I-T(N)]\}^{-1}  \label{Eq10}
\end{equation}
the previous expression can be rewritten in the following compact form: 
\begin{equation}
S(N,N+1)=\Xi (N)\times S(N,N).  \label{Eq11}
\end{equation}
Inserting Eq. (\ref{Eq11}) into Eq. (\ref{Eq9}) we finally obtain: 
\begin{equation}
S(n,N+1)=S(n,N)+S(n,N)\times H\times \lbrack I-T(N)]\times \Xi (N)\times
S(N,N).  \label{Eq12}
\end{equation}
This equation still has an unknown quantity $S(N,N)$ which must be
determined separately. We achieve this combining the original transfer
matrix equation (\ref{Eq1}) and Eq. (\ref{Eq11}) to obtain the following: 
\begin{equation}
S(N+1,N+1)=T_{N}\times \Xi (N)\times S(N,N).  \label{Eq13}
\end{equation}
Eq. (\ref{Eq13}) is a nonlinear matrix equation with an initial condition
given by 
\begin{equation}
(G+H)\times S(0,0)=I.  \label{Eq14}
\end{equation}
Eqs. (\ref{tfromS})-(\ref{S_init}) of the main body of the paper coincide
with Eqs.(\ref{Eq12})-(\ref{Eq14}) with simplifyed notation for the matrix $%
S $, where we dropped the second argument. They constitute the complete set
of embedding equations for the transfer matrix problem. In order to find the
transmission coefficient one has to multiply the matrix $S(N,N)$ by the
boundary vector $v$; the first component of the resulting vector is equal to 
$t\exp (ikL)$, where $t$ is the complex transmission coefficient. If one is
interested in the distribution of the state vector $u(n,N)$ throughout the
entire system, one has to find $S(N,N)$ and then to solve Eq. (\ref{Eq12}).

The presented algorithm was proved to be extremely stable, it produced
reliable results for transmission as small as $10^{-17}$. This stability is
due to the operation of inversion involved in the calculations [see Eq. (\ref
{Eq10})]. This operation prevents elements of the matrix $S$ to grow
uncontrollably in the course of calculations.

\
\
\newpage 
\centerline{\bf Figure Captions} \noindent Fig. 1. Frequency dependence of
the averaged transmission coefficient for small concentrations of the
defects. The frequency is normalized by the fundamental ($k=0$) frequency of
the pure chain, $\Omega_0$. The low-frequency boundary of the polariton gap
is at $\omega \approx 1.3$ and is not shown here.

\noindent Fig. 2. The same as in Fig. 1 but for intermediate concentrations.

\noindent Fig. 3. The same as in Fig. 1 but for large concentrations.

\noindent Fig. 4. Concentration dependence of the collective localization
length, $l_{chain}$, normalized by the system's size $L$.

\noindent Fig. 5. Frequency dependence of the Lyapunov exponent of the entire
chain for several concentrations in the frequency region of the defect band.

\noindent Fig. 6. Concentration dependence of the semiwidth of the defect
band. The solid line represents fit with power function $c^{\nu}$, where $%
\nu \approx 0.8$.


\begin{references}
\bibitem{Yablonovitch2}  E. Yablonovitch, T.J. Gmitter, R.D. Meade, A.M.
Rappe, K.D. Brommer, and J.D. Joannopoulos. Phys. Rev. Lett. {\bf 67} 3380
(1991).

\bibitem{photonic}  J.D. Joannopoulos, R.D. Meade, J.N. Winn. {\it Photonic
Crystals: Molding the Flow of Light }(Princeton Univ. Press, 1995).

\bibitem{Yablonovitch1}  E. Yablonovitch, Phys. Rev. Lett. {\bf 58}, 2059
(1987).

\bibitem{Joannopoulos}  R.D. Meade, K.D. Brommer, A.M. Rappe, and J.D.
Joannopoulos, Phys. Rev. B {\bf 44} 13772 (1991).

\bibitem{Smith} D.R. Smith, R. Dalichaouch, N. Kroll, S. Schultz, S.L. McCall and P.M. Platzman, J. Opt. Soc. Am. B {\bf 10}, 314 (1993). 

\bibitem{Figotin}  A. Figotin and A. Klein, J. Stat. Phys. {\bf 86}, 165
(1997).

\bibitem{Sakoda} K. Sakoda and H. Shiroma, Phys. Rev. B {\bf 56}, 4830 (1997).

\bibitem{Deych1}  L.I. Deych and A.A. Lisyansky, Bull. Amer. Phys. Soc. {\bf %
42}, 203 (1997); Phys. Lett. A, {\bf 240}, 329 (1998).

\bibitem{Deych2}  L.I. Deych and A.A. Lisyansky, Phys. Lett. A, {\bf 243},
156 (1998).

\bibitem{Podolsky}  V.S. Podolsky, L.I. Deych, and A.A. Lisyansky, Phys.
Rev. B, {\bf 57}, 5168 (1998).

\bibitem{footnote} There is another type of local photon and polariton states, which must be
distinguished from those discussed in the paper. These are so called
photon-atom\cite{John1,John2,John3} and polariton-atom\cite{Rupasov} bound
states that arise in both photonic and polariton band-gaps due to optically
active atom impurities.

\bibitem{John1}  S. John and T. Quang, Phys. Rev. A {\bf 50}, 1764 (1994).

\bibitem{John2}  S. John and J. Wang, Phys. Rev. Lett. {\bf 64}, 2418
(1990); Phys. Rev. B {\bf 43}, 12772 (1991).

\bibitem{John3}  S. John and T. Quang, Phys. Rev. A, {\bf 52}, 4083 (1995).

\bibitem{Rupasov}  V.I. Rupasov and M. Singh, Phys. Rev. A, {\bf 54}, 3614
(1996); Phys. Rev. A {\bf 56}, 898 (1997).

\bibitem{electrontunneling}  A.V. Chaplik and M.V. Entin, Zh. Eksp. \& Teor.
Fiz. {\bf 67}, 208 (1974) [Sov. Phys.- JETP {\bf 40}]

\bibitem{Lifshitz}  I.M. Lifshitz and V. Ya Kirpichenkov, Zh. Eksp. \& Teor.
Fiz. {\bf 77}, 989 (1979) [Sov. Phys.- JETP {\bf 50}].

\bibitem{IEM}  R. Bellman and G. Wing, {\it An introduction to invariant
embedding} (Wiley, New York, 1976).

\bibitem{Penley}  J.C. Penley, Phys. Rev. {\bf 128}, 596 (1962).

\bibitem{Azbel}  B. Ricco and M.Ya. Azbel, Phys. Rev. B {\bf 29}, 1970
(1984).

\bibitem{Goldman-1}  V.J. Goldman, D.C. Tsui, and J.E. Cunningham, Phys.
Rev. Lett. {\bf 58}, 1256 (1987).

\bibitem{Goldman-2}  V.J. Goldman, D.C. Tsui, and J.E. Cunningham, Phys.
Rev. B. {\bf 35}, 9387 (1987);

\bibitem{Goldman-3}  A. Zaslavsky, V.J. Goldman, D.C. Tsui, and J.E.
Cunningham, Appl. Phys. Lett. {\bf 53}, 1408 (1988).


\end{references}
\end{document}